# Optical discrimination of *live single* cancer cells using reflection-based nanohole array sensor

Alfredo Franco[1,2, †], Izan Calderón[1, †], Dolores Ortiz[1,2], José L. Fernández-Luna[2,3], Fernando Moreno[1,2]

*Abstract*— In this research, a reflection-based nanohole array sensor system is presented for discriminating between migration-competent cancer cells that maintain the integrity of the actin cortex and those cells lacking the actin cortex and thus unable to migrate. Unlike previous transmission-based approaches, this configuration allows for more practical integration into *in situ* diagnostic tools. For the first time, the system performance is analyzed by studying the spectral features of the reflected light by *live single cells*. We demonstrate that the presence of the actin cortex, needed for cell migration, in different types of cancer cells significantly affect their optical response, enabling high sensitivity and specificity in cell classification. Our results pave the way for reflection-based plasmonic biosensor devices as a compact and efficient platform for developing biomedical application tools.

## I. INTRODUCTION

Nanohole arrays on metallic films have emerged as a powerful tool for plasmonic sensing due to Extraordinary Optical Transmission phenomenon (EOT), enabling high-sensitivity detection of biomolecular interactions [1,2]. EOT-based sensors leverage the excitation of surface plasmons in metallic nanostructures to detect variations in the refractive index (RI) of materials near the sensor surface. This property has been widely exploited in biomedical applications, including DNA and protein detection, molecular binding studies, and cellular response analysis [3,4]. Despite the success of EOT in detecting small amounts of biomolecules, its application to larger and more complex biological structures, such as whole cells, remains underexplored [5]. Current studies primarily focus on cell-substrate interactions, such as cell adhesion and morphology characterization [6], but cell optical properties can give some keys to differentiate different kinds of cells [7]. In particular, the ability to directly differentiate between cancer and non-cancer cells based on their intrinsic optical properties remains a challenge. Given that eukaryotic cells possess an outer membrane enriched with proteins and an actin cortex that plays a key role in the cellular mechanics, investigating their optical responses in relation to these structural components may provide new insights into cellular behavior. Usually, EOT-based biosensors work in transmission mode but depending on the application and the sensor configuration, it can operate in reflection mode which can be more suitable, for instance for designing some chirurgical tools [8]. Both transmission and reflection based EOT sensors have interesting advantages as compared with other kinds of sensors. They enable label-free performance, which eliminates the need for chemical markers or fluorescent dyes, preserving the natural state of biological samples and simplifying sample preparation. Compared to highly specific biosensors that require functionalization for detecting particular biomolecules, label-free EOT sensors provide a more general approach, allowing the detection of broad classes of biomaterials based on their optical properties [9].

In this work, we present the performance of a **reflection-based nanohole array sensor system** to differentiate control *live single cancer cells* from treated *live single cancer cells with disrupted actin cortex*, which impedes cell migration, a feature of aggressive cancer cells. This research complements that developed in [10] and paves the way for the development of tools for practical biomedical applications. By utilizing reflected light instead of transmitted signals, the proposed setup simplifies optical alignment, enhances signal collection efficiency, and facilitates integration into compact, fiber-optic based biosensing platforms. Notably, this method builds upon prior studies on transmission-based EOT sensors for cellular analysis [10, 11] and tissue differentiation [12] but offers advantages in terms of practical implementation and adaptability to point-of-care applications. We demonstrate for the first time the feasibility of the reflection approach by analyzing spectral shifts due to the change of the optical properties of cancer cells according to the organization of their actin cortex, highlighting its influence on their optical response and consequently, on the cell dynamics.

This work was supported by Ministerio de Ciencia e Innovación grant PID2021-128220NB-I00, the Instituto de Salud Carlos III through grant DTS18/00141, co-funded by the European Regional Development Fund/European Social Fund, "A way to make Europe/Investing in your future", and the Instituto de Investigación Valdecilla (IDIVAL) (APG/03). Corresponding author: Fernando Moreno (morenof@unican.es).

[1] *Department of Applied Physics, Faculty of Sciences, University of Cantabria, 39005 Santander, Spain*
[2] *Valdecilla Research Institute (IDIVAL), 39012 Santander, Spain*
[3] *Genetics Unit, Valdecilla University Hospital, 39008 Santander, Spain*
[†] *These authors contributed equally to this work.*

## II. Materials and methods

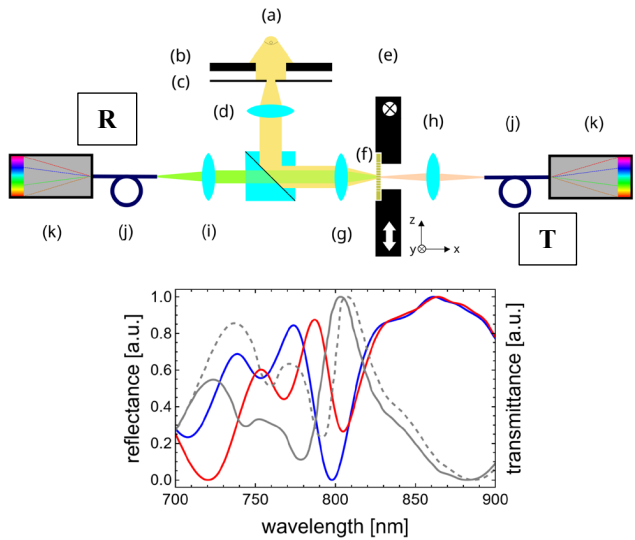

**Fig. 1.** *Top*: Experimental set-up scheme: (a) The light source is a 100 W halogen lamp; (b) Field stop: the minimum field stop diameter is 1.5 mm; (c) Field pinhole: it is a geometric tailorable pinhole; (d) The condenser is an achromatic and aplanatic lens; (e) The sample-holder is a stage for displacements along both x and y directions; (f) The nanohole patterned sensor; (g) The objective lens is a 20× bright field lens; (h,i) The collecting lens is an achromatic 4.5 mm focal distance lens; (j) The optical fiber (QP200-2-UV-VIS, Ocean Optics, USA) has a 200 μm core, optimized for its use in the UV–visible-NIR spectral region; (k) spectrograph (Andor Shamrock, Oxford Instruments, UK) operating with a 300 μm entrance slit and coupled to an CCD camera (Andor IDUS, Oxford Instruments, UK) with an integration time of 0.1 s. R/T refers to reflection/transmission configuration. *Bottom*: An example of the experimental results for the transmission (*Continuous grey line*: treated cells, *broken grey line*: control cells) and reflection (*blue line*: treated cells, *red line*: control cells) arrangements.

### A. Fabrication of Nanostructured Films

Gold thin films (80 nm) were deposited onto glass substrates and patterned with periodic (780 nm) subwavelength holes (340 nm of diameter), forming a 500μm x 500μm square array, via focused ion beam (FIB) milling (CIC nanogune (https://www.nanogune.eu/en), San Sebastian, Spain). The hole arrays were designed to support surface plasmon resonances (SPRs) in the far-red-NIR range.

### B. Cell Preparation and Measurement Protocol

Colorectal cancer cell lines HT29, CaCo2 and SW480 were from ATCC (Manassas, VA). Cells were grown as monolayers; HT29 cells were cultured in McCoy's 5A, CaCo2 in Dulbecco's Modified Eagle Medium (DMEM) (both from Biowest, Nuaillé, France) and SW480 in Leibovitz's L-15 medium (ATCC) each supplemented with 10% fetal bovine serum (GE Healthcare Hyclone, Chicago, IL), 100 U/mL penicillin and 100 μg/mL streptomycin (Lonza, Basel, Switzerland) and incubated in a humidified chamber at 37 °C and 5% $CO_2$. The medium was replaced every 2–3 days. At the time of measurement, cells were detached from the culture plate with a trypsin solution (Thermo Fisher Scientific, Waltham, MA), washed with phosphate-buffered saline (PBS) (Merck, Darmstadt, Germany) and aliquoted at a density of $5\times10^5$ cells/mL in PBS. HT29, CaCo2 and SW480 ($1\times10^6$ cells) were treated with Mycalolide B (SantaCruz Biotechnology, Dallas, TX) at a final concentration of 5μM for 1h at 37 °C and washed twice in PBS. Then, an aliquot ($5\times10^4$ cells) was used for optical measurements, and the rest of the cells were used for immunofluorescence staining. Cells staining was carried out incubating them with tetramethyl rhodamine conjugated phalloidin (Sigma-Aldrich) for 1 h at room temperature for visualization of F-actin, then they were washed in PBS and incubated with 0.2 μg/mL DAPI (Vector Lab, Burlingame, CA) for 10 min at room temperature to control viability. Following centrifugation, cells were resuspended in 15 μL of ProLong Gold Antifade Reagent (Thermo Fisher Scientific) and mounted onto microscope slides. Immunofluorescence images of the samples were obtained with an inverted fluorescence microscope (Axio Observer 5, Carl Zeiss Microscopy, Jena, Germany) using a 10x objective.

For optical measurements, 0,45 μl of the aliquot was deposited on the nanohole array and it was covered with a glass coverslip. The reflection spectra resulting from the interaction of single cells with the nanostructures were measured by means of the setup depicted below, restricting the illumination spot to 20μm. Measured spectra were smoothed by code to determine a clear peaks position. These spectra were compared to the PBS background spectrum to gather final spectra from which the refractive index within the plasmon skin is inferred. The measurements were carried out using three different cell lines to ensure the results are not cell-dependent, and all the measurements were performed three independent times to ensure results reproducibility.

### C. Reflection-Based Optical Setup

The system designed to measure the reflection spectrum resulting from the interaction between single cells and the nanoholes array is a modified version of an optical upright bright field microscope (Eclipse Ni, Nikon Instruments Inc., Japan), working either in transmission or reflection modes (Figure 1).

The system can measure the transmission/reflection spectrum in tailored regions as small as a few $\mu m^2$ in such a way that the measurement of a single cell can be isolated (see insets of Fig. 2). The diameter enclosing such regions is going to be denoted by D from now on. In this setup, the pupil of the microscope built-in field stop is modified by means of the incorporation of a tailored pinhole (denoted as *field pinhole* from now on), placed above the field stop. The condenser lens is shifted to a new position placed at a distance above its built-in position in such a way that the new distances from the nanohole patterned surface to the condensing lens and to the *field pinhole* produces a 0.1 lateral magnification. Consequently, if the light passes through a circular *field pinhole* of diameter D′, it is focused on the nanopatterned surface in a circular spot of diameter D=D′/10. The light transmitted/reflected through the illuminated area of the chip surface is collected by the objective lens and, by means of the collecting lens, it is focused on the entrance of the optical fiber to be transmitted to the spectrograph where the spectrum is acquired. To increase the signal-noise ratio, the spectra are the result of the accumulation of 180 consecutive spectral measurements. Restriction in the illumination area allows us to inspect the specific interaction between a single cell and the nanostructured surface with an increased sensitivity, because the transmission/reflection spectra are sensitive to the characteristics (optical, geometrical, etc.) of the whole illuminated region. The bottom part of Fig. 1 illustrates a typical result of the transmitted (grey lines)/reflected (red/blue lines) spectra (see Fig. 1 caption for details). Two main features can be seen: both spectra are complementary as expected [13] and they show a typical blue shift corresponding to a decrease of the local refractive index of *single* treated cells, in comparison to control cells, due to their disrupted actin cortex.

## III. RESULTS

The refractive index of both treated and control cells was determined from the position of the peak near 770 nm in their reflection spectra. For each single cell, the shift of this peak, relative to its position when the sample only consists of PBS, was used to calculate the refractive index within the plasmon penetration skin depth, considering the measured optical sensitivity of the nanohole array (402 nm/RIU).

Fig. 2 shows the histograms of the refractive index values obtained from measurements on both control (with actin cortex) and treated (without actin cortex) single HT29, CaCo2 and SW480 cells. The mean refractive index values are respectively, (1.334/1.337), (1.340/1.343) and (1.334/1.342) where the first value corresponds to the treated cells and the second to the control ones. In all cases, most of the measurements corresponding to control cells (red histogram) show higher refractive index values compared to those from treated cells (blue histograms). The presence of the actin cortex in the plasmon field penetration increases the value of the refractive index seen by the corresponding electromagnetic surface wave. The mean statistical sensitivities of the measurements are 0.85, 0.73 and 0.75, while the mean specificities are 0.64, 0.87 and 0.79, for HT29, CaCo2 and SW480 cells, respectively.

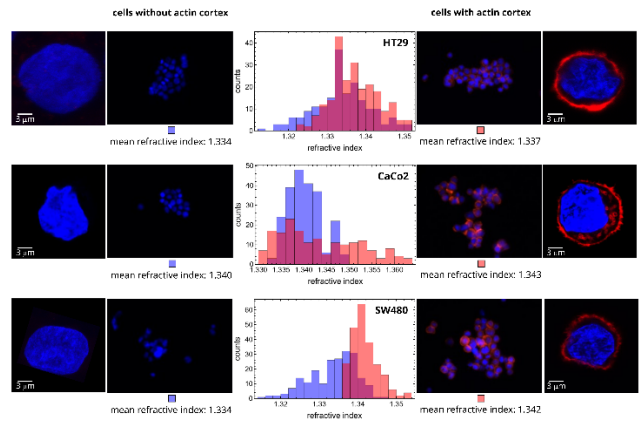

**Fig. 2.** Histograms of refractive index values for control and treated HT29, CaCo2 and SW480 cells. Red histograms correspond to control cells, while blue histograms represent treated cells. Insets show fluorescence microscopy images of representative control (right) and treated (left) cells. In the fluorescence images, blue indicates cell nuclei stained with DAPI, and red indicates the actin cortex labeled with fluorescent phalloidin.

Fluorescence microscopy images in Fig. 2 are representative examples of the measured cells. Both, an ensemble of analyzed cells and an isolated cell, as seen through our experimental set-up (Fig. 1), are shown for illustrative purposes. The images on the left show live treated cells, with their nuclei labeled with DAPI fluorescence (blue) and no signal from fluorescent phalloidin (red), indicative of actin cortex disruption. In contrast, the images on the right depict live control cells, which display both DAPI fluorescence and the characteristic red signal from phalloidin bound to the intact actin cortex.

## IV. DISCUSSION

Our study demonstrates for the first time the potential of a reflection-based biosensor for label-free discrimination between different types of *live single cells*. In contrast to traditional transmission-based systems, the reflection configuration offers several advantages for integration into compact, fiber-optic devices and in situ applications [6,9]. This design eliminates the need for a dual-port optical setup, reducing alignment complexity and potentially improving the signal-to-noise ratio through more efficient light collection.

One of the key advantages of a label-free sensor is its ability to detect intrinsic differences in cellular optical properties without the need for external markers or dyes. In our system, control cells from different lines exhibited a consistent redshift in the reflection spectrum compared to those with disrupted actin cortex. This redshift appears to be primarily correlated with differences in the actin cortex —a structural component critical for maintaining cell shape and mechanical integrity. The actin cortex is known to undergo significant remodeling during tumorigenesis, leading to variations in cellular stiffness and refractive index. Our observations, including reduced spectral separation upon actin disruption, further support the hypothesis that the organization of the actin network plays a major role in determining the effective refractive index sensed by the device. Furthermore, our reflection-based approach builds upon previous transmission-based EOT studies for cellular analysis [10-11]. Although transmission methods have successfully demonstrated high sensitivity for biomolecular detection, they often require more complex optical alignment and can be less robust in practical applications. In contrast, the reflection configuration simplifies the experimental setup and facilitates seamless integration with fiber-optic systems, making it well-suited for in situ and point-of-care diagnostic applications [8]. Despite these promising results, several challenges remain. Our current study was performed under controlled laboratory conditions with a limited number of cell types and samples. The reproducibility and long-term stability of the reflection-based sensor in complex biological environments need further investigation. Additionally, variations in ambient conditions and sample handling may influence the spectral response. Future work should focus on implementing rigorous calibration protocols, expanding the range of tested cell types, and integrating the sensor with microfluidic platforms to enhance sample delivery and measurement repeatability.

## V. Conclusion

Our reflection-based metallic nanohole biosensor represents a promising tool for cellular analysis, offering a simpler, more robust alternative to traditional transmission-based systems. As an example, its ability to differentiate *live single cancer cells* based on intrinsic optical properties of the actin cortex opens new avenues for non-invasive cancer diagnostics and the study of cellular biomechanics. In addition to further optimization and validation, reflected-based configurations of metallic nanohole biosensors can also be extended for developing point-of-care devices for tissue diagnostics [12].

## Acknowledgment

Authors want to be grateful to Olga Gutiérrez (Valdecilla Research Institute) for her technical assistance.

## References

1. T.W. Ebbesen, *et al*., "Extraordinary optical transmission through subwavelength hole arrays," Nature 391, 667–669 (1998).
2. H.F. Ghaemi, *et al*., "Surface plasmons enhance optical transmission through subwavelength holes," Phys. Rev. B 58, 6779–6782 (1998).
3. R. Gordon, *et al*., "A new generation of sensors based on extraordinary optical transmission," Acc. Chem. Res. 41(8), 1049–1057 (2008).
4. A.G. Brolo, *et al*., "Surface plasmon sensor based on the enhanced light transmission through arrays of nanoholes in gold films," Langmuir 20, 4813–4815 (2004).
5. N. Díaz-Herrera, *et al*., "Refractive index sensing of aqueous media based on plasmonic resonance in tapered optical fibers operating in the 1.5 μm region," Sens. Actuators B Chem. 146, 195–198 (2010).
6. J. Homola, "Surface plasmon resonance sensors for detection of chemical and biological species," *Chem. Rev.* 108, 462–493 (2008).
7. Liu, P. Y., *et al*. Cell refractive index for cell biology and disease diagnosis: past, present and future. *Lab on a Chip*, *16*(4), 634–644 (2016).
8. Lan, X., *et al*. Reflection based extraordinary optical transmission fiber optic probe for refractive index sensing. *Sensors and Actuators, B: Chemical*, *193*, 95–99 (2014).
9. T.-Y. Chang, *et al*., "Large-scale plasmonic microarrays for label-free high-throughput screening," *Lab Chip* 11, 3596–3602 (2011).
10. Carcelen, M., *et al*. Plasmonic Biosensing for Label-Free Detection of Two Hallmarks of Cancer Cells: Cell-Matrix Interaction and Cell Division. *Biosensors*, *12*(9), 674 (2022).
11. A. Franco, *et al*., "A nanohole array biosensor for discriminating live single cancer cells from normal cells," *Nanophotonics* 11(2), 315–328 (2022).
12. V. García-Milán *et al*. Discriminating Glioblastoma from Peritumoral Tissue by a Nanohole Array-Based Optical and Label-Free Biosensor. *Biosensors* 13(6), 591-603 (2023).
13. Søndergaard, T., *et al*. Extraordinary optical transmission enhanced by nanofocusing. *Nano Letters*, *10*(8), 3123–3128 (2010).